\documentclass[]{mn2e}

\usepackage{epsfig,rotate,graphicx}

\title{Eccentricity pumping of a
planet on an inclined orbit by a disc}

\author[Terquem and Ajmia]{Caroline Terquem$^{1,2}$\thanks{E-mail:
caroline.terquem@iap.fr} and Aikel Ajmia$^1$ \\ $^1$ Institut
d'Astrophysique de Paris, UPMC Univ Paris 06, CNRS, UMR7095, 98 bis bd
Arago, F-75014, Paris, France \\ $^2$ Institut Universitaire de France }

\begin{document}

\maketitle

\begin{abstract}
  In this paper, we show that the eccentricity of a planet on an
  inclined orbit with respect to a disc can be pumped up to high values
  by the gravitational potential of the disc, even when the orbit of the
  planet crosses the disc plane.  This process is an extension of the
  Kozai effect.  If the orbit of the planet is well inside the disc
  inner cavity, the process is formally identical to the classical Kozai
  effect.  If the planet's orbit crosses the disc but most of the disc
  mass is beyond the orbit, the eccentricity of the planet grows when
  the initial angle between the orbit and the disc is larger than some
  critical value which may be significantly smaller than the classical
  value of 39~degrees.  Both the eccentricity and the inclination angle
  then vary periodically with time.  When the period of the oscillations
  of the eccentricity is smaller than the disc lifetime, the planet may
  be left on an eccentric orbit as the disc dissipates.
\end{abstract}


\section{Introduction}
\label{sec:intro}

Among the 240 extrasolar planets that have been detected so far with a
semi--major axis larger than 0.1 astronomical unit (au), about 100 have
an eccentricity $e > 0.3$.  Five of them even have $e>0.8$.  Such large
eccentricities, which cannot be the result of disc--planet interaction
(Papaloizou et al. 2001), are probably produced by planet--planet
interactions, either through scattering or secular perturbation (see
Ford \& Rasio~2008 and references therein), that occur after the disc
dissipates (Juric \& Tremaine 2008, Chatterjee et al. 2008, Ford \&
Rasio 2008).

Here, we show that high eccentricities can be pumped {\em by the disc}
if the orbit of the planet is inclined with respect to the disc.  The
process involved is an extension of the Kozai mechanism, in which a
planet is perturbed by a distant companion on an inclined orbit (Kozai
1962).  While the Kozai effect has always been studied for the case in
which the companion is far away from the planet, the process
investigated here is shown to be efficient even if the orbit of the
planet crosses the disc.  The classical Kozai effect has of course been
very well studied.  Here, we show that some significant differences
occur when the classical scenario is extended to apply to a disc.

In section~\ref{sec:Kozai} we review the Kozai effect, and show that the
same behaviour is expected whether the planet is perturbed by a distant
companion or by a ring of material orbiting far away.  In
section~\ref{sec:simulations}, we present the results of numerical
simulations of the interaction between a planet on an inclined orbit and
a disc.  We show that, provided most of the mass in the disc is beyond
the orbit, and the initial inclination is larger than some critical
value, the gravitational potential from the disc causes the eccentricity
and the inclination of the planet's orbit to oscillate with time.  This
may occur even if the orbit crosses the disc.  In
section~\ref{sec:discussion} we summarise our findings, and discuss
under which conditions this mechanism could operate.  The important
result is that a planet on an inclined orbit with respect to the disc
and located in or within the planet formation region may have its
eccentricity pumped up to high values by the interaction with the disc.
This is of astronomical interest, since inclinations are beginning to be
measured for extrasolar planets.

\section{Review of the Kozai effect and extension to a disc}
\label{sec:Kozai}

We consider a planet of mass $M_p$ orbiting around a star of mass
$M_\star$ which is itself surrounded by a ring of material of mass
$M_{\rm disc}$. The ring is in the equatorial plane of the star whereas
the orbit of the planet is inclined with respect to this plane.  The
motion of the planet is dominated by the star, so that its orbit is an
ellipse slightly perturbed by the gravitational potential of the ring.
We study the secular perturbation of the orbit due to the ring.  We
denote by $(X,Y,Z)$ the Cartesian coordinate system centred on the star
and $(r, \varphi, \theta)$ the associated spherical coordinates. The
ring is in the $(X,Y)$--plane between the radii $R_i$ and $R_o>R_i$.  We
suppose that the angular momentum of the disc is large compared to that
of the planet's orbit so that the effect of the planet on the disc is
negligible: the disc does not precess and its orientation is invariable.
The gravitational potential exerted by the ring at the location of the
planet is:

\begin{equation}
\Phi = -G \int_{R_i}^{R_o} \Sigma(r) r dr \int_0^{2 \pi}
\frac{d \alpha}{\left( r^2+r_p^2-2rr_p \cos \alpha \sin \theta_p \right)^{1/2}},
\label{Phi}
\end{equation}

\noindent where the subscript $p$ refers to the planet and $\Sigma(r)$
is the mass density in the ring.  We assume:

\begin{equation}
\Sigma(r)=\Sigma_0 \left( \frac{r}{R_o} \right)^{-n},
\label{sigma}
\end{equation}

\noindent where:

\begin{equation}
\Sigma_0 = \frac{(-n+2) M_{\rm disc}}{2 \left( 1-\eta^{-n+2} \right)\pi R_o^2 }, 
\label{sigma0}
\end{equation}

\noindent with $\eta \equiv R_i/R_o$.  We suppose that $R_i \gg r_p$,
so that the square root in equation~(\ref{Phi}) can be expanded in
$r_p/r$ and integrated to give:

\begin{eqnarray}
\Phi=- \frac{-n+2 }{1-\eta^{-n+2} } \frac{G M_{\rm
 disc}}{R_o}
 \left[    \frac{1 - \eta^{1-n}}{1-n}  +   \; \; \; \; \; \; \; \; \; \;
  \; \; \; \; \; \right. \\
 \left.  \; \; \; \; \; \; \; \; \; \; \; \; \; \; \; 
\frac{ -1 + \eta^{-1-n} }{ 1+n } \frac{r_p^2}{2R_o^2}
\left( -1 + \frac{3}{2} \sin^2 \theta_p \right)
\right].
\end{eqnarray}

In the classical Kozai effect, the planet is perturbed by a distant
companion of mass $M$.  If we assume the orbit of this outer companion
is circular of radius $R \gg r_p$ and lies in the $(X,Y)$--plane, then
the potential averaged over time it exerts at the location $(r_p,
\theta_p)$ is:

\begin{equation}
\Phi_{\rm Kozai} = -\frac{GM}{R} \left[ 1+ \frac{r_p^2}{2R^2} 
\left( -1 + \frac{3}{2} \sin^2 \theta_p \right)  \right] .
\end{equation}

Because $r_p$ and $\theta_p$ appears in exactly the same way in $\Phi$
and $\Phi_{\rm Kozai}$, the secular perturbation on the inner planet,
obtained by averaging over the mean anomaly of its orbit, is the same in
both cases to within an overall multiplicative factor.  The results
obtained for the classical Kozai effect can therefore be extended to the
case of the disc.  In particular, the perturbation due to the disc makes
the eccentricity $e$ of the planet to oscillate with time if the initial
inclination angle $I_0$ between the orbit of the planet and the plane of
the disc is larger than a critical angle $I_c$ given by:

\begin{equation} 
\cos^2 I_c = \frac{3}{5}.  
\end{equation}

\noindent The maximum value reached by the eccentricity is then (Innanen
et al.~1997):

\begin{equation}
e_{\rm max}=\left( 1- \frac{5}{3} \cos^2 I_0 \right)^{1/2},
\label{emax}
\end{equation}

\noindent and the time $t_{\rm evol}$ it takes to reach $e_{\rm max}$
starting from $e_0$ is (Innanen et al.~1997): 

\begin{equation}
\frac{t_{\rm evol}}{\tau} =  
0.42  
\left( \sin^2 I_0 - \frac{2}{5}  \right)^{-1/2} 
\ln \left( \frac{e_{\rm max}}{e_0}  \right),
\label{tevol}
\end{equation}

\noindent with the time $\tau$ defined as:

\begin{equation}
\tau=\frac{(1+n)(1-\eta^{-n+2})}{(-n+2)(-1+\eta^{-n-1})}
 \frac{R_o^3 M_{\star}}{a^3 M_{\rm disc}} \frac{T}{2 \pi},
\label{tau}
\end{equation}

\noindent where $T$ is the orbital period of the planet and $a$ is its
semi--major axis.  Note that the function $ \left( \sin^2 I_0 - 2/5
\right)^{-1/2}$ decreases very sharply from infinity to $\sim 3$ as
$I_0$ increases from $I_c$ to about 45$^{\circ}$ and then decreases by
about 50\% as $I_0$ continues to increase up to 90$^{\circ}$.

The $Z$--component of the angular momentum of the orbit, $L_z \propto
\sqrt{1-e^2} \cos I$, where $I$ is the inclination angle between the
orbit and the plane of the disc, is constant.  Therefore $I$ also
oscillates with time and is out of phase with $e$.

\section{Numerical simulations}
\label{sec:simulations}

We consider a star of mass $M_{\star}=1$~M$_{\sun}$ surrounded by a disc
of mass $M_{\rm disc}$ and a planet of mass $M_p$ whose orbit is
inclined with respect to the disc.  The planet interacts gravitationally
with the star and the disc but we take $M_p \ll M_{\rm disc}$ so that it
has no effect on the disc.  To study the evolution of the system, we use
the $N$--body code described in Papaloizou~\& Terquem~(2001) in which we
have added the gravitational force exerted by the disc onto the planet.

The equation of motion for the planet is:
\begin{equation} 
{d^2 {\bf r} \over dt^2} = -{GM_\star{\bf r} \over |{\bf r}|^3} -
\mbox{\boldmath $\nabla$} \Phi - {GM_p{\bf r} \over |{\bf r}|^3} + {\bf
\Gamma}_{t,r} \; , \label{emot}
\end{equation} 

\noindent where ${\bf r}$ is the position vector of the planet and
$\Phi$ is the gravitational potential of the disc given by
equation~(\ref{Phi}) with $R_i$ and $R_o$ being the inner and outer
radii of the disc.  The third term on the right--hand side is the
acceleration of the coordinate system based on the central star.  Tides
raised by the star in the planet and relativistic effects are included
through ${\bf \Gamma}_{t,r}$, but they are unimportant here as the
planet does not approach the star closely.  Equation~(\ref{emot}) is
integrated using the Bulirsch--Stoer method and the integrals involved
in $\mbox{\boldmath $\nabla$} \Phi$ are calculated with the Romberg
method (Press et al. 1993).  In most runs, the integration conserves
the total energy of the planet and $L_Z$ within 1 to 2\%.

The planet is set on a circular orbit at the distance $r_p$ from the
star.  The initial inclination angle of the orbit with respect to the
disc is $I_0$.  In the simulations reported here we have taken $n=1/2$
in equation~(\ref{sigma}).  The functional form of $\Sigma$ is shallower
than what is usually used for discs, but that has no significant effect
on the argument we develop here.

\begin{figure}
  \includegraphics[width=14cm]{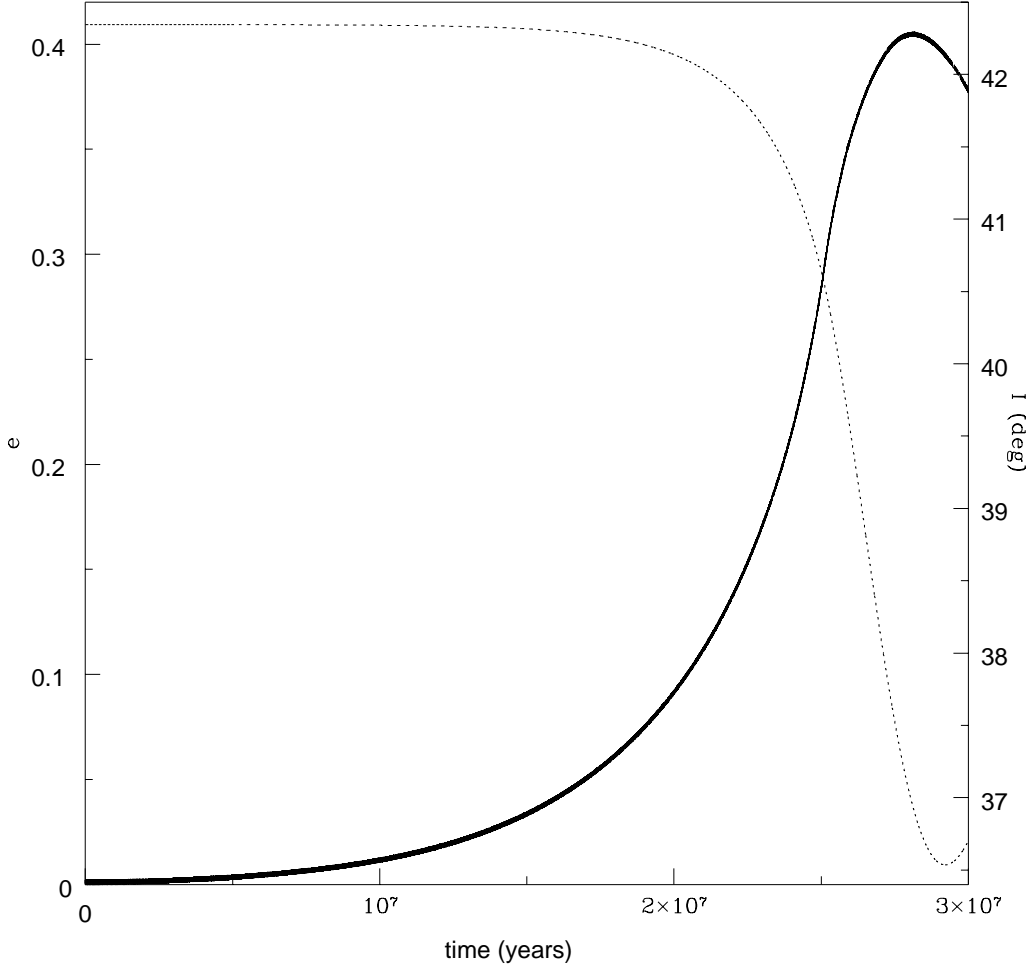} 
  \caption{Eccentricity $e$ ({\em solid line})
    and inclination angle $I$ (in degrees, {\em dotted line}) versus
    time (in years) for $M_p=10^{-3}$~M$_{\sun}$, $r_p=1$~au,
    $M_{\rm disc}=10^{-2}$~M$_{\sun}$, $R_o=100$~au, $R_i=50$~au and
    $I_0=42^{\circ}.3$.} 
     \label{fig1}
\end{figure}

We first compare the numerical results with the analysis summarised in
section~\ref{sec:Kozai} by setting up a case with $R_i \gg r_p$.  In
figure~\ref{fig1} we display the evolution of $e$ and $I$ for
$M_p=10^{-3}$~M$_{\sun}$, $r_p=1$~au, $M_{\rm disc}=10^{-2}$~M$_{\sun}$,
$R_o=100$~au, $R_i=50$~au and $I_0=42^{\circ}.3$.  For this run, $L_Z$
is conserved within 2\% but the energy of the planet is conserved only
within 10\%.  We are here in the conditions of the analysis of
section~\ref{sec:Kozai} with $\eta=0.5$.  From equation~(\ref{emax}), we
expect $e_{\rm max}=0.3$, which is a bit smaller than the value of 0.41
found in the simulation.  Also the minimum value of $I$ should be
$I_c=39^{\circ}.2$ and is observed to be $36^{\circ}.5$.  Note that
since the energy varies by about 10\% in this run, we do not expect
exact agreement between the numerical and the analytical results.  We
observe that the time it takes to reach $e_{\rm max}$ from the initial
conditions is $2.8 \times 10^7$~years, which agrees well with $t_{\rm
evol}$ given by equation~(\ref{tevol}) provided we take $e_0 \simeq 2
\times 10^{-2}$.  As mentioned above, $t_{\rm evol}$ becomes very long
when $I_0$ is smaller than 45$^{\circ}$.  As the disc lifetime is only a
few Myr, $e$ would not have time to reach the maximum value in this
case, if starting from a very small value.

\begin{figure}
  \includegraphics[width=14cm]{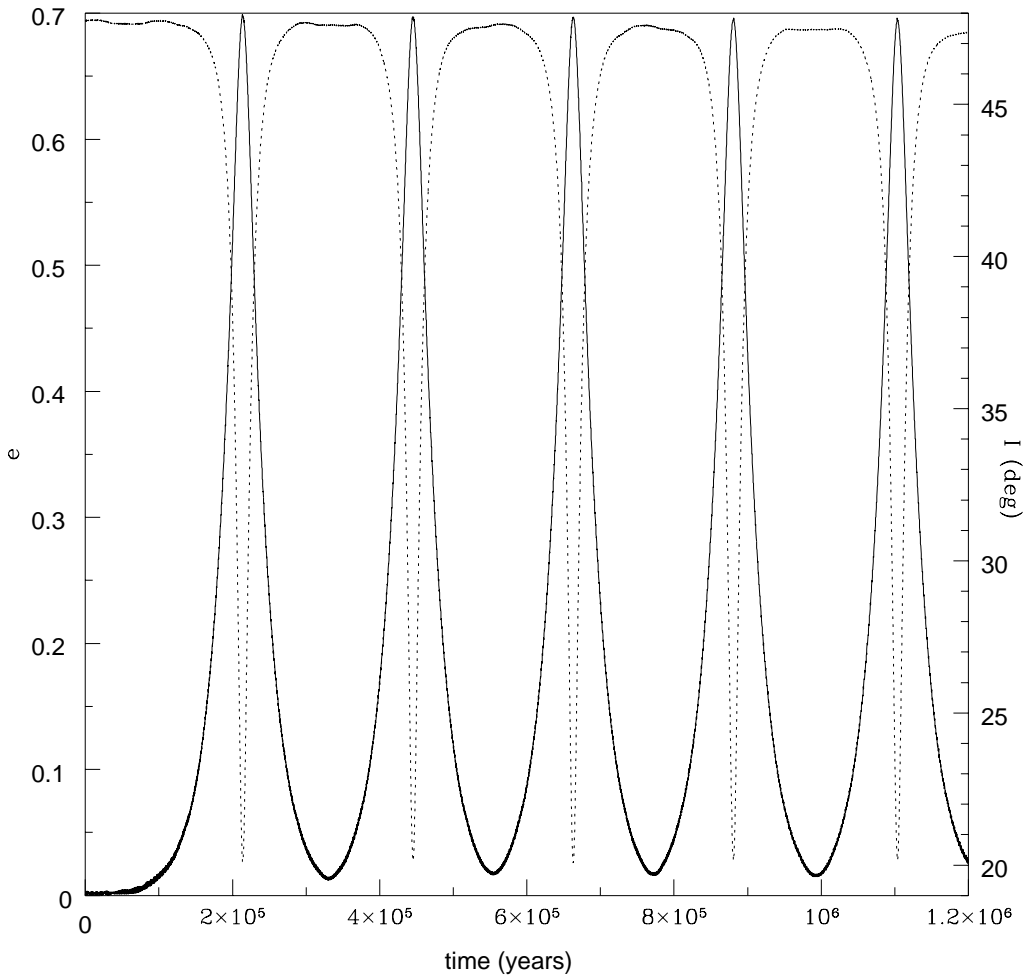} 
  \caption{ Same as figure~\ref{fig1} but for
   $r_p=20$~au, $R_i=1$~au and $I_0=47^{\circ}.7$.}  
   \label{fig3}
\end{figure}

Figure~\ref{fig3} shows the evolution of $e$ and $I$ for
$M_p=10^{-3}$~M$_{\sun}$, $r_p=20$~au, $M_{\rm
disc}=10^{-2}$~M$_{\sun}$, $R_o=100$~au, $R_i=1$~au and
$I_0=47^{\circ}.7$ (case~A).  We see that $e$ oscillates between $e_{\rm
min}=10^{-2}$ and $e_{\rm max}=0.7$, whereas $I$ oscillates between
$I_{\rm min}=20^{\circ}.1$ and $I_{\rm max}=I_0$.  The values of $e_{\rm
min}$, $e_{\rm max}$ and $I_{\rm min}$ differ from those calculated in
the analysis in section~\ref{sec:Kozai}, but this is expected as the
condition $r_p \ll R_i$, that was used in the analysis, is not valid
here.  However, since most of the mass in the disc is in the outer
parts, beyond the planet's orbit, the behaviour we get here is similar to
that described in the analysis.  The period of the oscillations is
$T_{\rm osc}=2.2 \times 10^5$~years.

\begin{figure}
  \includegraphics[width=14cm]{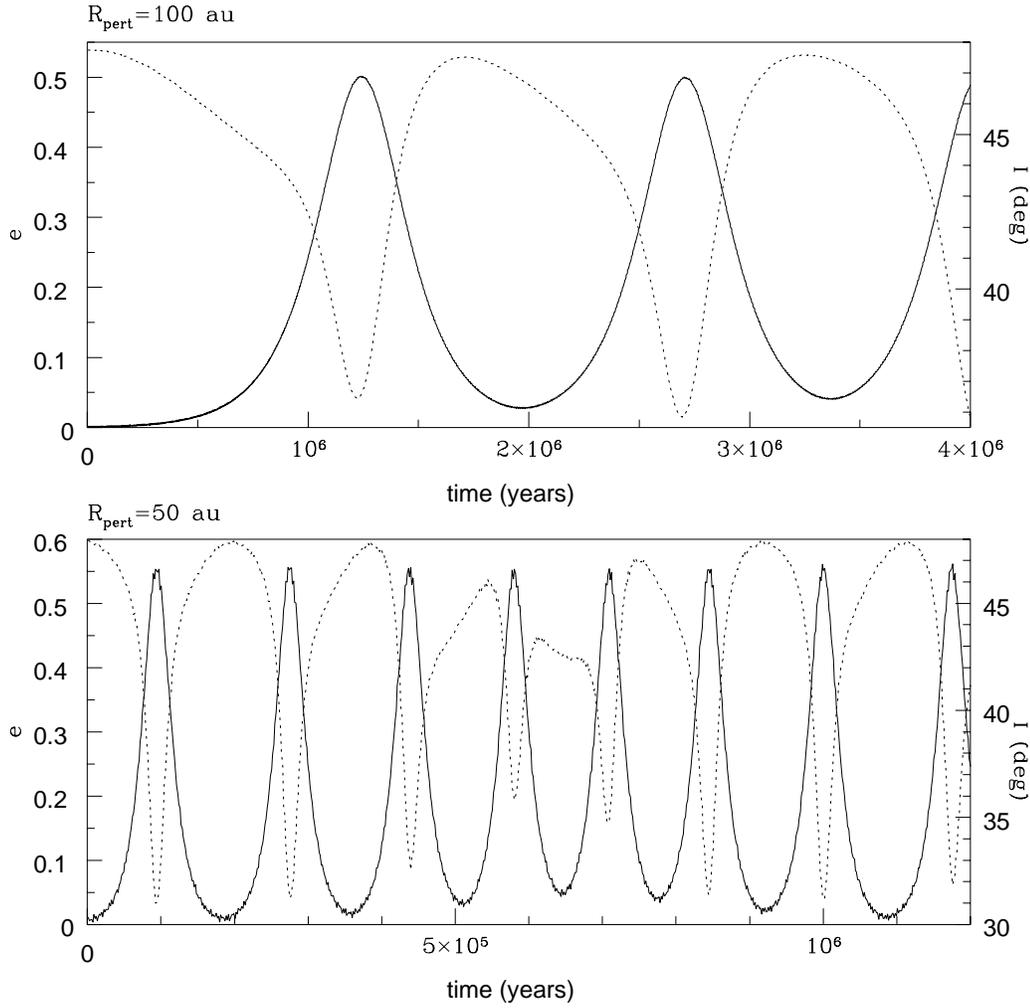} \caption{Classical
 Kozai effect: Same as figure~\ref{fig3} but with the disc being
 replaced by a planet of mass $M=10^{-2}$~M$_{\sun}$ and located at a
 distance $R=100$~au ({\em upper plot}) and 50~au ({\em lower plot})
 from the star.}  \label{fig3bis}
\end{figure}

For comparison, we have run the classical Kozai case, where the disc is
replaced by a planet located on a circular orbit in the $(X,Y)$--plane.
This perturbing planet is at a distance $R$ from the central star and
has a mass $M$.  We take $M$ to be the same as the value of $M_{\rm
disc}$ above, and consider $R=50$ and 100~au.  The evolution of $e$ and
$I$ for the inner planet in that case is shown in
Figure~\ref{fig3bis}. Equation~(\ref{emax}) gives $e_{\rm max}=0.5$,
which is in very good agreement with the values of 0.5 and 0.55 obtained
from the numerical simulations for $R=100$ and 50~au, respectively.  The
time it takes to reach $e_{\rm max}$ from $e_{\rm min}$, which is
$T_{\rm osc}/2$, is given by equation~(\ref{tevol}) with $e_0$ being
replaced by $e_{\rm min}$ and $\tau=[R^3 M_{\star} / (a^3 M)] T / (2
\pi)$.  We get $t_{\rm evol} = 5.5 \times 10^5 $ and $8 \times
10^4$~years for $R=100$~au ($e_{\rm min}=0.03$) and 50~au ($e_{\rm
min}=0.02$), respectively, which is in excellent agreement with the
values seen on Figure~\ref{fig3bis}.  The minimum angle reached when
$R=50$~au is about $31^{\circ}$, much larger than the value obtained
when the disc is present.  Of course this value would decrease if the
perturbing planet were moved closer to the inner planet, but then the
oscillations would not be regular anymore, as can already be seen in the
case $R=50$~au.

\begin{figure}
  \includegraphics[width=14cm]{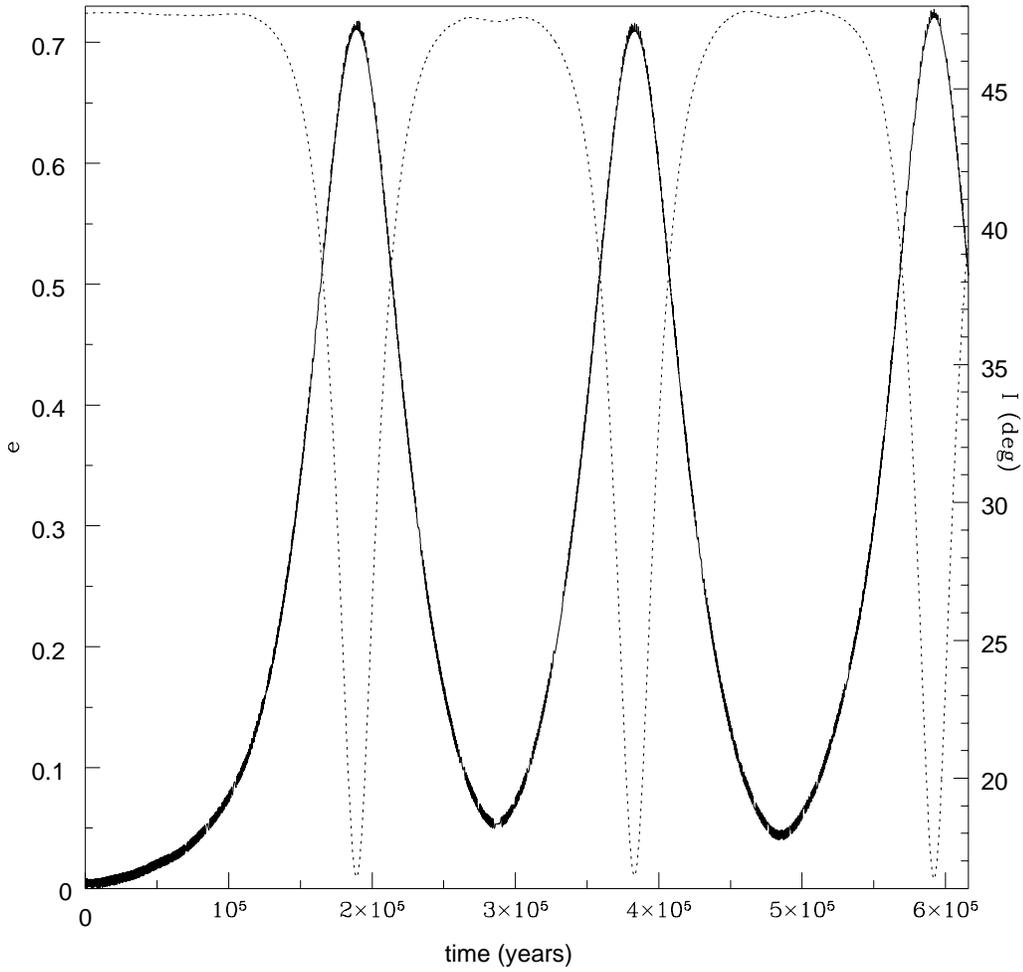} 
  \caption{Same as figure~\ref{fig3} but for
    $M_p=4 \times 10^{-3}$~M$_{\sun}$, $r_p=1$~au, $R_o=50$~au,
    $R_i=0.5$~au.}  
    \label{fig4}
\end{figure}

Figure~\ref{fig4} shows the evolution of $e$ and $I$ for the same values
of $M_{\rm disc}$ and $I_0$ as in case~A but with $M_p=4 \times
10^{-3}$~M$_{\sun}$, $r_p=1$~au, $R_o=50$~au and $R_i=0.5$~au (case~B).
Here we have $e_{\rm min}=5 \times 10^{-2}$, $e_{\rm max}=0.72$, $I_{\rm
min}=16^{\circ}.4$, $I_{\rm max}=I_0$ and $T_{\rm osc} = 2.0 \times
10^5$~years, comparable to the value found in the previous case.

On dimensional grounds, $T_{\rm osc}$ is expected to be proportional to
$\tau$ given by equation~(\ref{tau}).  We have run case~A with different
values of $M_{\rm disc}$ and have checked that $T_{\rm osc} \propto
1/M_{\rm disc}$.  The values between which $e$ and $I$ oscillate though
do not depend on $M_{\rm disc}$, as expected from the analysis.


We have run case~A with different values of $r_p$ ranging from 2 to
50~au.  We observe that for $r_p$ roughly below 10~au, $T_{\rm osc}$
decreases when $r_p$ increases, as expected from the expression of
$\tau$.  For larger values of $r_p$ though, $T_{\rm osc}$ does not vary
much with $r_p$, and is $ \sim 2$--$3 \times 10^5$~years.  For $r_p=10$
and 20~au, the extreme values of $e$ and $I$ are roughly the same if the
calculations are started from the same $I_0$ and $e_0$, but for
$r_p=50$~au the amplitude of the oscillations is very small.  In that
case, most of the disc mass if not beyond the planet's orbit, so that
the Kozai effect disappears.

Finally, we have checked the effect of varying $I_0$ in case~A.  We have
found that there is a critical value of $I_0$, $I_c \sim 30^{\circ}$,
below which eccentricity growth was not observed.  However, the fact
that $I_{\rm min}=20^{\circ}.1$ in case~A suggests that $I_c$ may
actually be smaller.  Since the time it takes for $e$ to grow from very
small values when $I_0$ is close to $I_c$ is very long (see
equation~[\ref{tevol}]), the simulations may not have been run long
enough for a growth to be observed.  As $I_0$ increases from $I_c$ to
90$^{\circ}$, $e_{\rm max}$ grows from 0 to 1.  $T_{\rm osc}$ is not
observed to vary significantly with $I_0$, although the time it takes
for the eccentricity to grow from its initial value to $e_{\rm max}$
does depend on $I_0$.

The simulations reported here suggest that when the planet's orbit
crosses the disc, eccentricity growth occurs for significantly smaller
initial inclination angles than in the classical Kozai effect.

\section{Discussion and conclusion}
\label{sec:discussion}

We have shown that, when a planet's orbit is inclined with respect to a
disc, the gravitational perturbation due to the disc results in the
eccentricity and the inclination of the orbit oscillating if: (i) most
of the disc mass is beyond the planet's orbit, (ii) the initial
inclination angle $I_0$ is larger than some critical value $I_c$, which
may be significantly smaller than in the classical Kozai effect.  In the
simulations we have performed, in which $\Sigma \propto r^{-1/2}$ and
$M_{\rm disc} \ll M_p$ (so that the effect of the planet on the disc is
negligible), oscillations occur as long as the initial planet's distance
to the star, $r_p$, is smaller than about half the disc radius $R_o$.
We expect a smaller critical distance when $\Sigma$ decreases more
rapidly with radius.  Note that $I_c$ should be independent of the
functional form of $\Sigma$ as long as most of the disc mass is beyond
the planet's orbit.  The amplitude of the oscillations depends only on
$I_0$.  It is small for $I_0 \simeq I_c$ and becomes large when $I_0$ is
increased.  In particular, $e_{\rm max} \rightarrow 1 $ as $I_0
\rightarrow 90^{\circ}$.  The oscillations of $e$ and $I$ are
$180^{\circ}$ out of phase.  Their period $T_{\rm osc} \propto 1/M_{\rm
disc}$.  When $r_p \ll R_o$, $T_{\rm osc}$ decreases as $r_p$ increases.
For larger values of $r_p$, $T_{\rm osc}$ does not vary much with this
parameter.  In the simulations we have performed, $T_{\rm osc} \sim
10^5$ years.  As this is much shorter than the disc lifetime, there is a
nonzero probability that the planet is left on a highly eccentric orbit
as the disc dissipates.

Note that the process discussed in this paper is not a mere trivial
extension of the Kozai effect.  Indeed, in the classical Kozai effect,
the periodic behaviour of $e$ is obtained because the dependence of the
perturbing gravitational potential on the planet's argument of
pericentre $\omega$ is through a term proportional to $\cos 2 \omega$.
When the orbit of the planet crosses the disc, the gravitational
potential has a very different form, and the behaviour of $e$ is much
more difficult to predict.

The effect discussed here could be inhibited if other processes induced
a precession of the planet's orbit on timescales shorter than $T_{\rm
osc}$.  That may happen if other planets are present in the system, or
because of dissipative tidal torques exerted by the disc, which have
been ignored here (Lubow \& Ogilvie 2001).  In the latter case, the
Kozai effect would then work only for planets orbiting inside the disc
inner cavity.

When the planet crosses the disc, loss of energy and angular momentum
circularises the orbit and aligns it with the disc plane on a timescale
$\sim T M_p/(\Sigma(r_p) R_p^2)$, where $R_p$ is the planet radius (Syer
et al. 1991, Ivanov et al. 1999), which can be smaller than $T_{\rm
osc}$ for $r_p$ larger than a few~au.  Taking this process into account
together with the Kozai effect may lead to equilibrium values of $e$ and
$I$.  This will be studied in a forthcoming paper.  Note that the usual
type~II migration mechanism that applies to planets orbiting in discs
would not be relevant here, as it happens when the planet orbits in a
gap that is locked in the disc evolution.  In the coplanar case, as the
disc spirals toward the central star, it carries along the gap and the
planet.  When the planet is on an inclined orbit, it may still open up a
gap but it is not locked in it, so that it is not being pushed in as the
disc spirals in.

The mechanism described here relies on the planet being on an inclined
orbit, which could happen as a result of: (i) dynamical relaxation of a
population of planets formed through fragmentation of a protostellar
envelope around a star surrounded by a disc (Papaloizou \&
Terquem~2001); (ii) mean motion resonances (Thommes \& Lissauer 2003,
also Yu \& Tremaine 2001) and (iii) gravitational interactions between
embryos during the planet formation stage (Levison et al. 1998,
Cresswell \& Nelson 2008).  In all cases, the process that makes the
orbits inclined also makes them eccentric.  According to the results
presented here, the disc could pump the eccentricities up to even larger
values.

Measures of the projected angle between the axes of the planet's orbit and the
stellar rotation, using the Rossiter--McLaughlin effect, are becoming
available.  So far, only the system XO--3, which has a $\sim 12$ Jupiter
masses planet on a 3.19 days orbit with $e=0.26$, has been shown to have a
spin--orbit misalignment of at least 37$^{\circ}$.3 (H\'ebrard et al. 2008,
Winn et al. 2009).  Misalignment has also been reported for the system
HD~80606, which has a $\sim 4$ Jupiter mass planet on a 111.44 days orbit
with $e= 0.93$ (Moutou et al. 2009, Pont et al. 2009). As HD80606 is a
component of a binary system, the classical Kozai effect could be responsible
for the misalignment in this system.  

We thank J. Papaloizou and S. Balbus for very useful
comments which have greatly improved the original version of this paper.
C.T. thanks the MPIA in Heidelberg, where part of this paper was
written, for hospitality and support.

\clearpage

\onecolumn

\end{document}